\documentclass[final]{IEEEtran}

\usepackage{amsthm,amssymb,graphicx,multirow,amsmath,color,amsfonts}
\usepackage[update,prepend]{epstopdf}
\usepackage[noadjust]{cite}
\usepackage[latin1]{inputenc}
\usepackage{tikz}
\usetikzlibrary{arrows,calc}		
\usepackage{bbm} 
\usepackage{pdfpages}
\usepackage{tabulary}
\usepackage{multirow}
\usepackage{comment}
\def\matern{{Mat{\'e}rn\;}}
\newcommand{\myeq}[1]{\mathrel{\overset{\makebox[0.07pt]{\mbox{(#1)}}}{=}}}
\newcommand{\myle}[1]{\mathrel{\overset{\makebox[0.07pt]{\mbox{(#1)}}}{\le}}}
\newcommand{\myge}[1]{\mathrel{\overset{\makebox[0.07pt]{\mbox{(#1)}}}{\ge}}}

\def\nb0{{\mathbf{0}}}
\def\nb1{{\mathbf{1}}}

\def\nbbR{{\mathbb{R}}}

\newtheorem{thm}{Theorem}

\newtheorem{prop}{Proposition}

\newtheorem{remark}{Remark}

\def\figref#1{Fig.\,\ref{#1}}%
\def\E{\mathbb{E}}

\def\P{\mathbb{P}}

\def\R{\mathbb{R}}

\allowdisplaybreaks 
\usepackage{setspace}	

\begin{document}
\graphicspath{{./Figures/}}
\title{
Nearest-Neighbor and Contact Distance
Distributions for \matern Cluster Process}
\author{Mehrnaz Afshang, Chiranjib Saha, and Harpreet S. Dhillon
\thanks{The authors are with Wireless@VT, Department of ECE, Virgina Tech, Blacksburg, VA, USA. Email: \{mehrnaz, csaha,  hdhillon\}@vt.edu. The support of the US NSF (Grants CCF-1464293, CNS-1617896) is   gratefully acknowledged.}  \vspace{-.4em}
}


\maketitle
\thispagestyle{empty}
\pagestyle{empty}
\begin{abstract} 
 In this letter, we derive the cumulative density function (CDF) of the nearest neighbor and contact distance distributions of the \matern cluster process (MCP) in $\mathbb{R}^2$. These results will be useful in the  performance analysis  of many real-world wireless networks that exhibit inter-node attraction. Using these results, we concretely demonstrate that the contact distance of the MCP stochastically dominates its nearest-neighbor distance as well as the contact distance of the homogeneous Poisson point process (PPP) with the same density.%
\end{abstract}
\begin{IEEEkeywords}
Stochastic geometry, \matern cluster process,
contact distance, and nearest-neighbor distance.
\end{IEEEkeywords}
\section{Introduction} \label{sec:intro}

Owing to its simplicity and tractability, Poisson cluster process (PCP) is a natural first choice for modeling point patterns that exhibit inter-point attraction. Such point patterns appear in many branches of science, such as ecology, cosmology, geodesy, seismology, and wireless networks. PCP acquires its tractability from its underlying homogeneous PPP which models the cluster centers (called {\em parent points}). 
Clusters of {\em offspring points} are then placed independently around each parent point to form a PCP \cite{haenggi2012stochastic}. 
Within each cluster, the offspring points are located independently of each other with the same distribution (with respect to the parent point). One can configure the density of the parent points and the distribution of offspring points to match a particular dataset.%

Wireless nodes exhibit some degree of spatial clustering in many real-world deployments~\cite{ganti2009interference,SahAfshDhiUnifiedHetNet2017,afshang2015fundamentals}. For instance, this clustering is quite prominent in the locations of cellular users due to the formation of {\em user hotspots}, which further leads to clustering of small cells deployed within these hotspots~\cite{SahAfshDhiUnifiedHetNet2017,afshang2015fundamentals}. While designing and analyzing wireless networks, such spatial clusters are often modeled as circular regions in which wireless nodes are assumed to be uniformly distributed. If we use such clusters in a PCP (users distributed uniformly at random in disks around parent points), the resulting process is called MCP. Naturally, MCP is of special interest to wireless networking community. In fact, it has recently been shown that one can use MCP to closely emulate several user and base station configurations used by 3rd Generation Partnership Project (3GPP) to model heterogeneous cellular networks~\cite{SahAfshDhiUnifiedHetNet2017}. Despite its obvious importance in modeling wireless networks, its applicability has remained limited due to the absence of easy-to-use characterizations of its basic properties, such as the contact and the nearest-neighbor distance distributions. Such distributions form the basis of stochastic analyses of these networks and will be the main focus of this letter.

Not surprisingly, the general formalism to compute distance distributions of PCPs is known in the probability literature~\cite{huber2009likelihood}. More specifically,~\cite{huber2009likelihood} derives the likelihood function of a general  multidimensional PCP using which one can, in principle, obtain nearest-neighbor and contact distance distributions for the MCP in $\nbbR^2$. However, this formalism is heavily measure-theoretic and the resulting expressions are unwieldy involving integrals over general (typically irregular) regions of $\nbbR^2$. As a result, a common way-out for any MCP-based analysis of wireless networks is to resort to the first-order statistic approximation of MCP by the homogeneous PPP with the same density as that of MCP and use the corresponding distributions of PPP, which are available in closed-form~\cite{Wang2016,HetPCPGhrayeb2015}. However, given the different natures of MCP (in general any PCP) and PPP, such simplifications may lead to deceptive insights. Motivated by such shortcomings, explicit expressions for the nearest-neighbor and contact distance distributions for a Thomas cluster process (TCP) in $\nbbR^2$ were derived in~\cite{AfshSahDhi2016Contact}. Unfortunately, extending these results to the MCP is not straightforward. As discussed in the sequel, the main challenge is in accurately handling the finite support of the distribution of the offspring points around the parent points of the MCP. 

{\em Contributions.} The main contribution of this letter is the derivation of explicit expressions for the CDFs of contact and nearest-neighbor distances for the MCP. The resulting expressions are computationally efficient and can be readily used in the performance analyses of wireless networks modeled as MCPs. Using these results, we formally show that the contact distance of MCP stochastically dominates its nearest-neighbor distance as well as the contact distance of the PPP with the same density as that of the MCP. 

\section{\matern Cluster Process} \label{sec:SysMod}
\matern  cluster process is an isotropic, stationary PCP formed by offspring points whose locations around~the  parent points   are  independently and identically  distributed (i.i.d.)  with~PDF $f_{\bf S}({\bf s})=\begin{cases} \frac{1}{\pi r_{\rm d}^2}; & \|{\bf s}\| \le r_{\rm d},\\
0; &\text{otherwise},
\end{cases}.$
Thus, the offspring  points are uniformly distributed in a ball of radius $r_{\rm d}$ centered at ${\bf x}$ denoted by ${\bf b }({\bf x}, r_{\rm d})$, where ${\bf x}$ represents the location of a parent point. The parent points $\{{\bf x}\}$ form a homogeneous PPP $\Phi_{\rm p} \subset \R^2$ with density $\lambda_{\rm p}$.  The complete process $\Psi$ can be expressed as ${\bf x}+{\bf s}={\bf z}\in \Psi\equiv\cup_{{\bf x}\in \Phi_{\rm p}}\{{\bf x}+{\cal B}^{\bf x}\},$
where $\{{\bf s}\} \equiv {\cal B}^{\bf x}$ denotes the offspring point process, which is a finite point process independent of $\Phi_{\rm p}$. The elements in  $\{{\bf z}\}\subset{\bf x}+{\cal B}^{\bf x}\equiv {\cal N}^{\bf x}$ are conditionally i.i.d. with PDF $f_{\bf Z}({\bf z}|{\bf x})=f_{\bf S}({\bf z}-{\bf x})$. The number of points in ${\cal B}^{\bf x}$ (denoted by $|{\cal B}^{\bf x}|$) is Poisson~with~mean~$\bar{m}$.


{\bf Notation}: We use bold-style letters  ({\bf z})  to denote vector in $\R^2$, serif-style letters ($z$) to denote Euclidean norm, i.e., $z=\|{\bf z}\|$, and $o$ to denote origin.
\section{Distance Distributions}
\subsection{Contact distance distribution}
\begin{figure}
\centering
\includegraphics[width=.5\columnwidth]{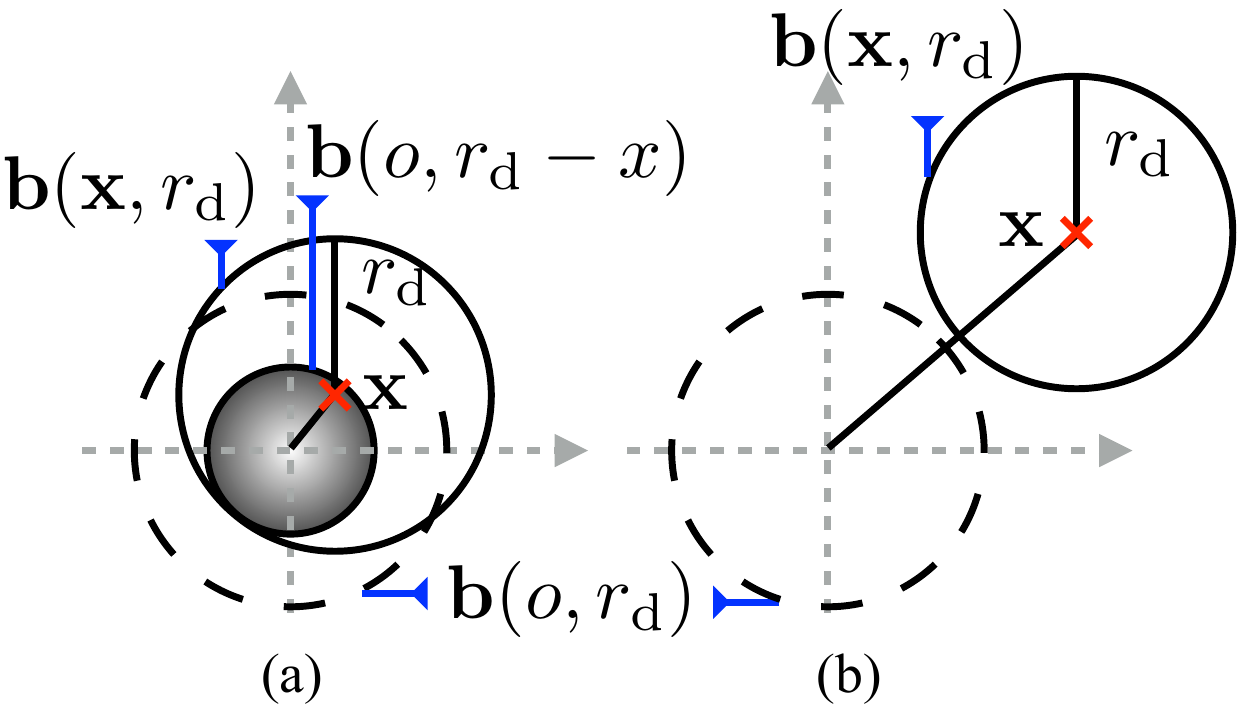}
\caption{ Possible position of cluster center from origin: (a) Case 1  (${\bf x}\in {\bf b}(o,r_{\rm d})$) and (b) Case 2 (${\bf x}\notin {\bf b}(o,r_{\rm d})$).}
\label{Fig Dis_cases}
\end{figure}

The contact distance  is the distance from a reference point (located at the origin)  to its nearest point of $\Psi$, where the reference point  is extraneous and independent  of the MCP ($o \notin \Psi$).
This is equivalent to the case where the reference point is sampled from a point process independent of the original  MCP.
The density function of  contact distance $R_{\rm C}$ (or, equivalently  empty space function) is defined as:
\begin{align}\label{eq: CDF Def contact distance}
F_{R_{\rm C}}(r)=1-\P(|\Psi({\bf b}( o,r))|=0),
\end{align}
where  $\Psi({\bf b}( o,r)) \equiv \Psi \cap {\bf b}( o,r)$.  Before deriving $F_{R_{\rm C}}(r)$, we first look at the distribution of distance from the reference point to an arbitrary point of   the set ${\cal N}^{\bf x}$ for a given ${\bf x}\in \Phi_{\rm p}$. Denote by  ${\cal D}^{\bf x}\equiv\{Z: Z=\|{\bf z}\|=\|{\bf x}+{\bf s}\|;\, \forall \, {\bf s} \in {\cal B}^{\bf x}\}$  the sequence  of  distances from elements of ${\cal N}^{\bf x}$ to the origin. 
In order to characterize the conditional PDF of $Z\in{\cal D}^{\bf x}$, it is required to consider the following two cases.\\
{\em Case 1:} The cluster centered at ${\bf x}\in \Phi_{\rm p}$ includes origin, i.e., ${\bf x}\in {\bf b}(o, r_{\rm d})$. For this case,  ${\bf z}\in \Psi$ can be either located within ${\cal A}_1\equiv {\bf b}(o, r_{\rm d}-x$) or ${\cal A}_2\equiv {\bf b}({\bf x}, r_{\rm d}) \setminus {\bf b}(o, r_{\rm d}-x)$; see \figref{Fig Dis_cases}.  For a given $x=\|{\bf x}\| <r_{\rm d}$, the elements in  ${\cal D}^{\bf x}$ are i.i.d. with conditional PDF~\cite[Theorem 2.3.6]{mathai1999introduction}
\begin{align} \label{Eq: Fu xi1 xi 2}
f_{Z}(z|{x})=\begin{cases}\chi^{(1)}(z,x);& 0 \le z\le r_{\rm d}-x,\\
\chi^{(2)}(z,{x}); &r_{\rm d}-x \le z \le r_{\rm d}+x,
\end{cases}
\end{align}
where~$
\chi^{(2)}(z,{x})=\frac{2 z}{\pi r_{\rm d}^2}\cos^{-1}\Big(\frac{z^2+x^2-r_{\rm d}^2}{2 z x}\Big)$~and~$\chi^{(1)}(z,{x})=\frac{2 z}{r_{\rm d}^2}$. \\
{\em Case 2:} The cluster centered at ${\bf x}\in \Phi_{\rm p}$ does not include origin, i.e., ${\bf x}\notin {\bf b}(o, r_{\rm d})$.  For this case, ${\bf z}\in \Psi$ is located in ${\cal A}_3= {\bf b}({\bf x}, r_{\rm d})$. For a given $x=\|{\bf x}\| >r_{\rm d}$, the elements in  ${\cal D}^{\bf x}$ are i.i.d. with conditional PDF~\cite[Theorem 2.3.6]{mathai1999introduction}
\begin{align}\label{eq chi3}
f_{Z}(z|{x})=\chi^{(3)}(z,{x})=\frac{2 z}{\pi r_{\rm d}^2}\cos^{-1}\Big(\frac{z^2+x^2-r_{\rm d}^2}{2 z x}\Big),
\end{align}
where $x-r_{\rm d}<z<x+r_{\rm d}$. The pictorial representation of these two cases is presented in \figref{Fig Dis_cases}. We now derive the CDF of contact distance, i.e., $R_{\rm C}$, in the next Theorem. The proof of this Theorem is presented in Appendix~\ref{proof of contact distance}.
\begin{thm} [Contact distance] The CDF of contact distance $F_{R_{\rm C}}(r)$ is given by~\eqref{eq: contact CDF}, at the top of next page.
\label{Thm: contact distance}
\end{thm}
\begin{figure*}
\begin{align}\notag
F_{R_{\rm C}}(r)&=1-\exp\Big(-2 \pi \lambda_{\rm p} \Big(  \int_0^{r_{\rm d}}  \Big(1- \exp\Big( -\bar{m} \Big( 
 \notag  \int_0^{\min(r, r_{\rm d}-x )} 
 \chi^{(1)}(z,x) {\rm d} z
 +  \int_{\min(r, r_{\rm d}-x )} ^{\min(r , r_{\rm d}+x)} \chi^{(2)}(z,x){\rm d} z   \Big)\Big) \Big) x  {\rm d} x \\ \label{eq: contact CDF}
& +\int_{r_{\rm d}}^{\infty}  \Big( 1-  \exp \Big(-\bar{m}   \int_{\min(r, x-r_{\rm d})}^{\min(r, x+r_{\rm d})}  \chi^{(3)}(z,x) {\rm d} z     \Big)\Big) x {\rm d} x \Big)         \Big)  ,
\end{align}
where $\chi^{(1)}(\cdot)$, $\chi^{(2)}(\cdot)$, and $\chi^{(3)}(\cdot)$ are given by  \eqref {Eq: Fu xi1 xi 2} and \eqref{eq chi3}. \\
\hrule
\end{figure*}
\begin{remark}
\label{Rem: challenge in MCP}
As observed from the proof of Theorem 1,  the main step in the derivation of above CDF is computing the probability that a cluster located at ${\bf x}$, i.e., ${\bf x}+{\cal B}^{\bf x}$, will have no point in ${\bf b}(o,r)$. Due to the finite support of each cluster in an MCP, this requires integrating $f_{\bf Z}({\bf z}|{\bf x})$ with respect to ${\bf z}$ over the region ${\bf b}(o,r)\cap {\bf b}({\bf x}, r_{\rm d})$  (see \eqref{eq proof contact}), which, depending on ${\bf x}$, assumes different shapes (circular or lens) as illustrated in \figref{Fig Dis_Contact}. On the contrary, for some PCPs with clusters that have infinite support, in particular TCP, the corresponding region is always ${\bf b}(o,r)$~\cite{AfshSahDhi2016Contact}. This makes such calculations more challenging for an MCP compared to say a TCP.  
\end{remark}


\subsection{Nearest-neighbor distance distribution}
\label{Subsec: Nearest-neighbor distance distribution}
The nearest-neighbor distance is  the distance of a reference point to its nearest point of $\Psi$, where  the reference point is a part of the original MCP (i.e., $o \in \Psi$).  The CDF of the nearest-neighbor distance, $R_{\rm N}$, can be  defined as:
\begin{align}
F_{R_{\rm N}}(r)&=1-\P(|\Psi({\bf b}( o,r))|=1 |  o \in \Psi)\\
&=1-\P^{!}_ o(|\Psi({\bf b}( o,r))|=0 ), \label{eq: NN definition}
\end{align}
where $\P^{!}_ o$ denotes reduced palm distribution. 
The nearest-neighbor distance can be equivalently defined as the contact distance when the  reference point is a point of $\Psi$ chosen at random with equal probability.  Denote by ${\bf x}_0 \in \Phi_{\rm p}$ the  center of  the reference point's own cluster. The probability mass function (PMF) of  the  number of offspring points within the reference point's own cluster, i.e., $|{\cal B}^{{\bf x}_0}|$,  is
\begin{align}\label{eq: number of points PMF V1}
\P(|{\cal B}^{{\bf x}_0}|=\ell)=\frac{\ell}{\bar{m}}\frac{\bar{m}^\ell e^{-\bar{m}}}{\ell ! } \quad \text{for } {\ell \in \mathbb{Z}^{+}}\: ,
\end{align} 
 where $\mathbb{Z}^{+}$ is set of positive integer. Note that $\P(|{\cal B}^{{\bf x}_0}|=0)=0$ because the reference point's own cluster cannot be empty.
The PMF of  $|{\cal B}^{{\bf x}_0}|$  is {\em number-weighted} and different from that of the number of points of a {\em typical} cluster  $|{\cal B}^{\bf x}|$ of $\Psi$, which is Poisson~\cite[Sec. 5.3]{chiu2013stochastic}. This  is because
when the reference point is selected uniformly at random, it is more likely to belong to the cluster of $\Psi$ with higher number of points (similar to the {\em waiting bus paradox}). Hence the selection of a cluster of $\Psi$ is biased by the number of its offspring points.
\begin{remark}
It is worth noting that if we add a point to the {\em typical} cluster ${\cal B}^{\bf x}$ (where the typical cluster is the one that is chosen uniformly at random from amongst all the clusters), the PMF of $|{\cal B}^{\bf x}|+1$ is the same as that of $|{\cal B}^{{\bf x}_0}|$. This property is a consequence of the fact that $|{\cal B}^{\bf x}|$ is Poisson distributed.
\end{remark}
In contrast to the PPP, where the contact  and nearest-neighbor distributions are identical, these two distributions are not the same for MCP. 
This  is because the selection of the reference point from MCP implies the existence of  a cluster ${\cal N}^{{\bf x}_0}\equiv \{{\bf x}_0+{\bf s}: {\bf s} \in {\cal B}^{{\bf x}_0}\}$ which includes origin, and hence $x_0=\|{\bf x}_0\|<r_{\rm d}$. Now, the CDF of nearest-neighbor distance can be expressed as
\begin{align}\notag
&F_{R_{\rm N}}(r)=1-\E\Big[\prod_{{\bf z}\in \Psi\setminus \{{o}\}} {\bf 1}\{{\bf z} \notin {\bf b}(o,r)\}\Big]\\\notag
&=1-\E\Big[\prod_{{\bf z}\in \Psi\setminus {\cal N}^{{\bf x}_0}} {\bf 1}\{{\bf z} \notin {\bf b}(o,r)\}\prod_{{\bf z}\in {{\cal N}^{{\bf x}_0} }\setminus \{o\}}{\bf 1}\{{\bf z} \notin {\bf b}(o,r)\}\Big]\\\notag
&\myeq{a}1-\E\Big[\prod_{{\bf z}\in \Psi} {\bf 1}\{{\bf z} \notin {\bf b}(o,r) \}\Big] \E \Big[\prod_{{\bf z}\in {{\cal N}^{{\bf x}_0}} \setminus \{o\}}{\bf 1}\{{\bf z} \notin {\bf b}(o,r)\}\Big]\\  \label{eq: proof of NN}
&\myeq{b}1- (1- F_{R_{\rm C}}(r))\E \Big[\prod_{{\bf z}\in {{\cal N}^{{\bf x}_0}} \setminus \{o\}}{\bf 1}\{{\bf z} \notin {\bf b}(o,r)\}\Big],
\end{align}
where ${\bf 1}\{\cdot\}$ denotes  indicator function. Step  (a) follows from Slivnyak's  Theorem which states that $\Psi$ equals to  $\Psi \setminus {\cal N}^{{\bf x}_0}$ in distribution, and $F_{R_{\rm C}}(\cdot)$ in (b) is given by~\eqref{eq: contact CDF}. From this step the final expression for the CDF of nearest-neighbor distance is presented in the next Theorem. The complete derivation of this result is presented in Appendix~\ref{App: Proof of NN distribution}.
\begin{thm}[Nearest-neighbor distance] \label{Thm: NN distance APP1}
The CDF of $R_{\rm N}$ is 
\begin{multline}
{F}_{R_{\rm N}}(r)=1-(1-F_{R_{\rm C}}(r))  \int_0^{r_{\rm d}}  \exp\Big( -\bar{m} \mu(x_0,r)   \Big) \\ \times f_{X_0}(x_0)  {\rm d} x_0 ,
\end{multline}
\begin{multline}
\text{with} \quad \mu(x_0,r)=\int_0^{\min(r, r_{\rm d}-x_0 )} 
 \chi^{(1)}(z,x_0) {\rm d} z \\
 +  \int_{\min(r, r_{\rm d}-x_0 )} ^{\min(r , r_{\rm d}+x_0)} \chi^{(2)}(z,x_0){\rm d} z, 
\end{multline}
 where   $\chi^{(1)}(\cdot)$,  $\chi^{(2)}(\cdot)$,    $F_{R_{\rm C}}(\cdot)$ are given respectively by  \eqref{Eq: Fu xi1 xi 2}, \eqref{eq: contact CDF}, and  $ f_{X_0}(x_0)=\frac{2 x_0}{r_{\rm d}^2} {\bf 1}(x_0\le r_{\rm d}) $.
\end{thm}
 \subsection{Establishing Stochastic Dominance}
 In the  next Proposition, we  show that the contact distance of MCP stochastically dominates the i) nearest-neighbor distance of MCP and ii) contact (or equivalently nearest-neighbor) distance of PPP with density matched to that of the MCP, which is $\bar{m}\lambda_{\rm p}$. The proof is provided in Appendix~\ref{App: proof stochastic dominance}.
 \begin{prop}\label{prop: contact distance}
Let $R_0$ be the contact distance of PPP with density $\bar{m} \lambda_{\rm p}$. Then
\begin{align*}
  R_{\rm C}  \ge_{\rm st}  R_{\rm N}\: \text{and}\:  R_{\rm C}  \ge_{\rm st} R_0,
\end{align*}
where $\ge_{\rm st}$ denotes first order stochastic dominance. Equivalently $F_{R_{\rm N}}(r) \ge F_{R_{\rm C}}(r)  \: \text{and}\: 1-\exp(\lambda_{\rm p} \bar{m} \pi r^2 ) \ge F_{R_{\rm C}}(r)$. 
 \end{prop}
 \begin{figure}
\centering
\includegraphics[width=.65\columnwidth]{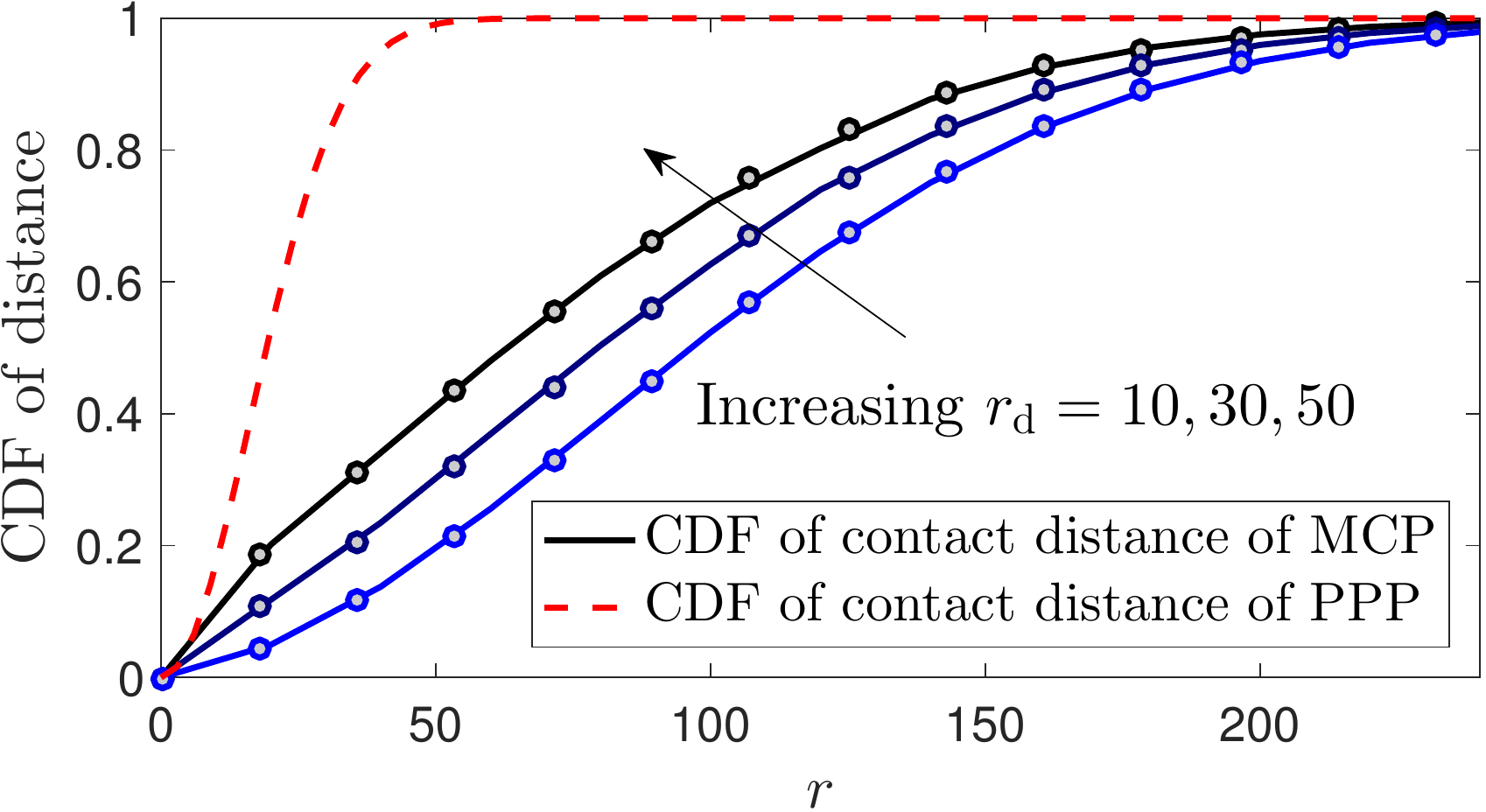}
\includegraphics[width=.65\columnwidth]{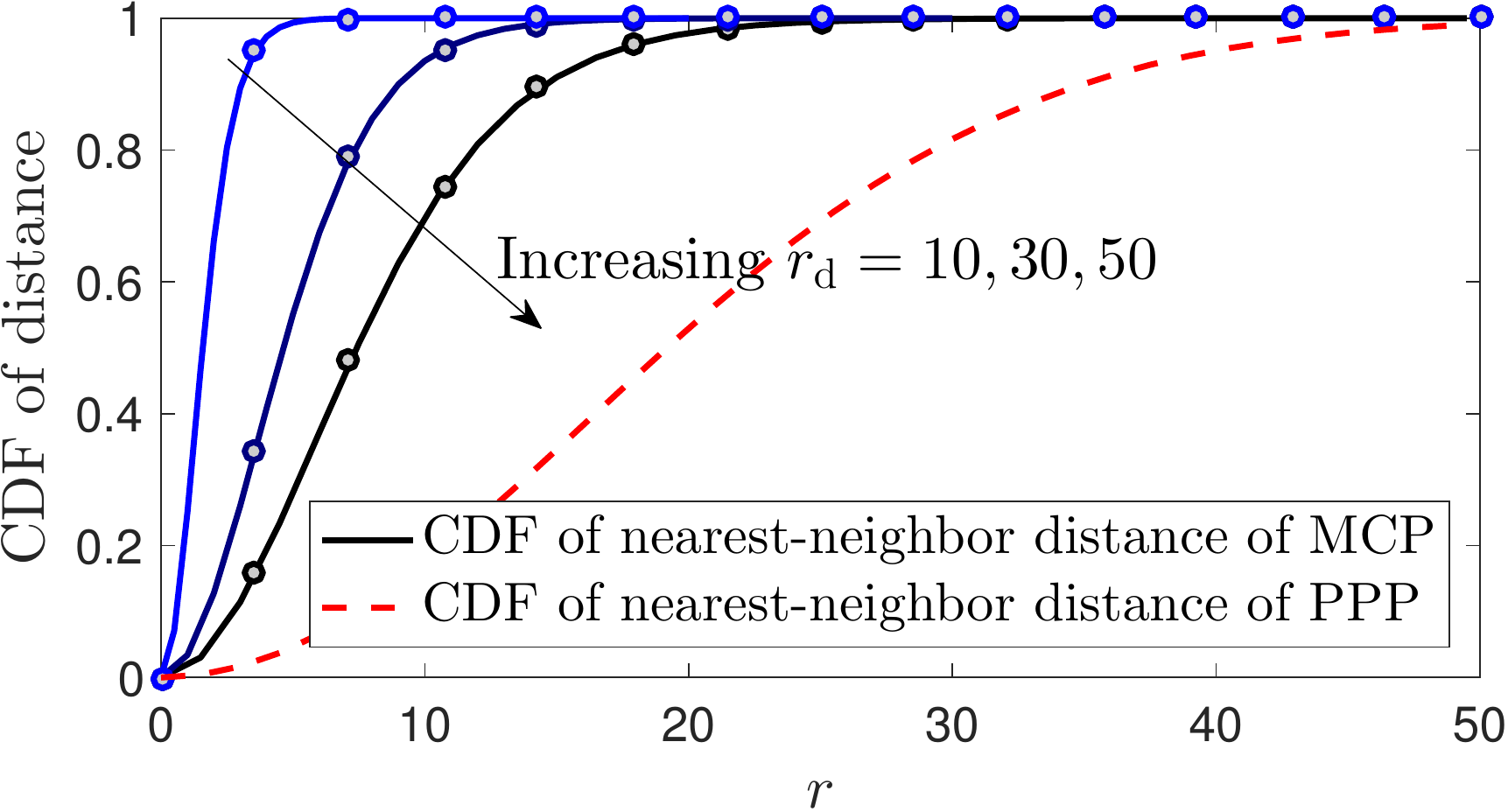}
\caption{CDFs of contact distance and nearest-neighbor distances ($\bar{m}=30, \lambda_{\rm p}=20\times 10^{-6} $). The markers correspond to simulation
results.}
\label{Fig DisMatern}
\end{figure}
In \figref{Fig DisMatern}, we plot CDFs of contact distance and nearest-neighbor distances. The perfect match between theoretical and simulation results corroborates the accuracy of our analysis. From these results, we notice that as we increase the cluster radius $r_{\rm d}$, the contact distance decreases while the nearest-neighbor distance increases. This is because when we increase $r_{\rm d}$ while keeping the number of points per cluster the same (on average), the clusters tend to become sparser, which increases inter-point distances within the cluster (which impacts the nearest-neighbor distance) and decreases the distance between an independent reference point and the closest point of the MCP (which is basically the contact distance). Both the CDFs converge to those for the PPP with density $\bar{m} \lambda_{\rm p}$ when $r_{\rm d}\to \infty$. This is consistent with the fact that PCP converges to a PPP as the {\em cluster sizes} are increased to infinity \cite{SahAfshDhiUnifiedHetNet2017}. 
\section{Concluding Remarks}
In this paper, we characterized the contact and nearest-neighbor distributions of the MCP.  We also demonstrated that the contact distance of the MCP stochastically dominates: i) its the nearest-neighbor distance, and ii) the contact distance of the PPP with the same density as that of the MCP.   Given the relevance of MCP in modeling real-world networks, these distributions enable the accurate characterization of useful  quantities of interest   such as received power in downlink, transmit power in uplink and key performance metrics such as  coverage probability, and area spectral efficiency. However, while the CDFs of both the contact and nearest-neighbor distances can be computed easily using standard numerical softwares, the presence of integrals in their expressions may lead to unwieldy expressions for network-wide performance metrics, such as coverage probability and area spectral efficiency. Consequently, a useful direction of future research is to derive easy-to-use, ideally closed-form, bounds or approximations for these expressions, which can be directly plugged into the analysis of key performance metrics of interest. 


\appendix
\subsection{Proof of  Theorem~\ref{Thm: contact distance}}
\label{proof of contact distance} 

\begin{figure}
\centering
\includegraphics[width=.8\columnwidth]{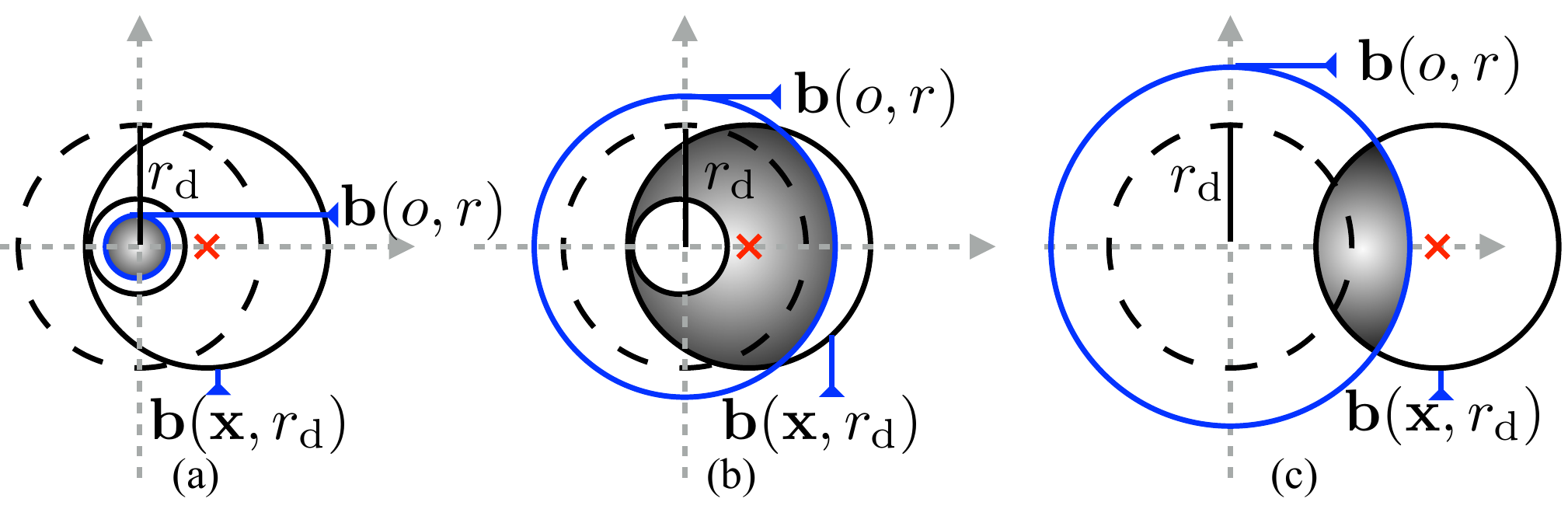}
\caption{ The shaded region illustrates ${{\bf b}({\bf x},r_{\rm d})\cap {\bf b}(o, r)}$.}
\label{Fig Dis_Contact}
\end{figure}

The  CDF of contact distance defined in~\eqref{eq: CDF Def contact distance}~can be written as $F_{R_{\rm C}}(r)=1-\P(|\Psi({\bf b}( o,r))|=0),
$
where $\P(|\Psi({\bf b}( o,r))|=0)$ is equal to $\E\Big[\prod_{\bf x \in \Phi_{\rm p}} \prod_{{\bf s} \in {\cal B}^{\bf x}} {\bf 1}\{({\bf x}+{\bf s}) \notin {\bf b}(o, r)\}\Big]$
\begin{align}\notag
 \myeq{a}&\exp\Big(-\lambda_{\rm p} \int_{\R^2}\Big(1- \exp\Big(-\bar{m} \\ \notag
 \times&  \int_{{\bf b}({\bf x},r_{\rm d})}  \Big(1- {\bf 1}\{({\bf x}+{\bf s}) \notin {\bf b}(o, r)\}\Big)  f_{\bf S}({\bf s}) {\rm d} {\bf s}\Big)\Big) {\rm d} {\bf x}\Big)\\ \notag
\myeq{b}&\exp\Big(-\lambda_{\rm p} \int_{\R^2}\Big(1- \exp\Big(-\bar{m} \\ \label{eq proof contact}
\times&  \int_{{\bf b}({\bf x},r_{\rm d})\cap {\bf b}(o, r)}  f_{\bf Z}({\bf z}|{\bf x}) {\rm d} {\bf z}\Big)\Big){\rm d} {\bf x}\Big).
\end{align}
Here (a) follows from the PGFL of a Poisson cluster process~\cite[Corollary 4.13]{haenggi2012stochastic}, and (b) follows from change of variable ${\bf x}+{\bf s}\to {\bf z}$, where  $f_{\bf Z}({\bf z}|{\bf x})=f_{\bf S}({\bf z}-{\bf x})$.
The inner integral in (b) can be simplified as follows.
 \begin{enumerate}
 \item  If $\|{\bf x}\|=x<r_{\rm d}$ and $r< r_{\rm d}-x$, then
 $$ \int_{{\bf b}({\bf x},r_{\rm d})\cap {\bf b}(o, r)}  f_{\bf Z}({\bf z}|{\bf x}) {\rm d} {\bf z}=\int_0^{\min(r, r_{\rm d}-x)} \chi^{(1)}(z,{x})  {\rm d} z,$$
where ${{\bf b}({\bf x},r_{\rm d})\cap {\bf b}(o, r)}$ is depicted in~\figref{Fig Dis_Contact}.a.
 \item If $x<r_{\rm d}$ and $ r_{\rm d}-x<r<r_{\rm d}+x$, then
  $$ \int_{{\bf b}({\bf x},r_{\rm d})\cap {\bf b}(o, r)}  f_{\bf Z}({\bf z}|{\bf x}) {\rm d} {\bf z}=\int_{\min(r, r_{\rm d}-x)}^{\min(r, r_{\rm d}+x)} \chi^{(2)}(z,{x})  {\rm d} z,$$
  where ${{\bf b}({\bf x},r_{\rm d})\cap {\bf b}(o, r)}$ is depicted in~\figref{Fig Dis_Contact}.b.
   \item If $x>r_{\rm d}$ and $ x-r_{\rm d}<r<x+r_{\rm d}$, then
  $$ \int_{{\bf b}({\bf x},r_{\rm d})\cap {\bf b}(o, r)}  f_{\bf Z}({\bf z}|{\bf x}) {\rm d} {\bf z}=\int_{\min(r, r_{\rm d}-x)}^{\min(r, r_{\rm d}+x)} \chi^{(3)}(z,{x})  {\rm d} z,$$
  where ${{\bf b}({\bf x},r_{\rm d})\cap {\bf b}(o, r)}$ is depicted in~\figref{Fig Dis_Contact}.c.
 \end{enumerate}
The final result is obtained by combining the above cases.
\subsection{Proof of  Theorem~\ref{Thm: NN distance APP1}}
\label{App: Proof of NN distribution}
Conditioning on ${\bf x}_0$, the expression $\E \big[\prod_{{\bf z}\in {{\cal N}^{{\bf x}_0}} \setminus \{o\}}{\bf 1}\{{\bf z} \notin {\bf b}(o,r)\}|{\bf x}_0\big]$  given in the step (b) of~\eqref{eq: proof of NN} can be simplified to
{\small
\begin{align*}
&{\sum_{\ell=1:\infty}  \Big(   \int_{{\bf b}({\bf x}_0,r_{\rm d})}  {\bf 1}\{{\bf z} \notin {\bf b}(o, r)\} f_{\bf Z}({\bf z}|{\bf x}_0) {\rm d} {\bf z}\Big)^{\ell-1} \frac{\ell}{\bar{m}}\frac{\bar{m}^\ell e^{-\bar{m}}}{\ell ! } }\\
&\myeq{c}\exp \Big(-\bar{m} \int_{{\bf b}({\bf x}_0,r_{\rm d}) \cap {\bf b}(o, r) }      f_{\bf Z}({\bf z}|{\bf x}_0) {\rm d} {\bf z}       \Big).
\end{align*}}
Given the fact that $o \in {{\cal N}^{{\bf x}_0}} \subset \Psi$, we have $\|{\bf x}_0\|< r_{\rm d}$. Thus, the inner integral in (c) can be simplified as follows.
 \begin{enumerate}
 \item If $ r <r_{\rm d}-x_0$, then
 $$ \int_{{\bf b}({\bf x}_0,r_{\rm d}) \cap {\bf b}(o, r) }      f_{\bf Z}({\bf z}|{\bf x}_0) {\rm d} {\bf z} =\int_0^{\min(r, r_{\rm d}-x_0)} \chi^{(1)}(z,{x}_0)  {\rm d} z. $$
 \item If $ r_{\rm d}-x_0<r<r_{\rm d}+x_0$, then
   $$ \int_{{\bf b}({\bf x}_0,r_{\rm d})\cap {\bf b}(o, r)}  f_{\bf Z}({\bf z}|{\bf x}_0) {\rm d} {\bf z}=\int_{\min(r, r_{\rm d}-x_0)}^{\min(r, r_{\rm d}+x_0)} \chi^{(2)}(z,{x}_0)  {\rm d} z.$$
 \end{enumerate}
Combining the above results and de-conditioning with respect to distance of the reference point to its cluster center $X_0$ with PDF $f_{X_0}(x_0)=\frac{2 x_0}{r_{\rm d}^2} {\bf 1}(x_0\le r_{\rm d})$,  we get the final result.
\subsection{Proof of Proposition~\ref{prop: contact distance}} 
\label{App: proof stochastic dominance}
Note that $R_{\rm C}  \ge_{\rm st}  R_{\rm N}$ if and only if the complementary CDF (CCDF) of $R_{\rm C}$ dominates that of $R_{\rm N}$. From  \eqref{eq: NN definition}, the CCDF  of $R_{\rm N}$ can be written as: $\P^{!}_ o(|\Psi({\bf b}( o,r))|=0 )$
\begin{align*}
 &\myeq{a}  \P(|\Psi({\bf b}( o,r))|=0 ) \P^{!}_ o(|{\cal N}^{{\bf x}_0}({\bf b}( o,r))|=0 )\\
&\myle{b} \P(|\Psi({\bf b}( o,r))|=0 ),
\end{align*}
where (a) follows from step (a) in \eqref{eq: proof of NN}, and (b) follows from the fact that  $\P^{!}_ o(|{\cal N}^{{\bf x}_0}({\bf b}( o,r))|=0 ) \in [0, 1]$.  Also, $\P(|\Psi({\bf b}( o,r))|=0 ) $ is equal to
\begin{align*}
& \exp\Big(- \lambda_{\rm p}\int_{\R^2} \Big(1-   \exp \Big(-\bar{m}   \int_{{\bf b}(o, r)} f_{\bf Z}( {\bf z}|{\bf x}) {\rm d} {\bf z} \Big)\Big){\rm d} {\bf x} \Big)\\
\myge{c}&   \exp\Big(- \lambda_{\rm p} \bar{m} \int_{\R^2} \int_{{\bf b}(o, r)}  f_{\bf Z}( {\bf z}|{\bf x}) {\rm d} {\bf z} {\rm d}{\bf x} \Big)\\
=&  \exp\Big(- \lambda_{\rm p} \bar{m}  \int_{{\bf b}(o, r)} \underbrace{ \int_{\R^2}  \frac{{\bf 1}\{\|{\bf z}-{\bf x}\| \le r_{\rm d}\}}{\pi r_{\rm d}^2}  {\rm d}{\bf x}}_{1} {\rm d} {\bf z}\Big)\\
=&  \exp\big(- \lambda_{\rm p} \bar{m} \pi r^2\big),
\end{align*}
where $(c)$ follows from $1-\exp(-\xi v)\le \xi v$, $\xi\ge 0$.
%
%
%
%
%
%
\bibliographystyle{IEEEtran}

\bibliography{ArXive_Metern_contact_NN_v13.bbl}


\end{document}